# A Scalable, High-Efficiency, Low-Energy-Spread, Laser Wakefield Accelerator using a Tri-plateau Plasma Channel

Shuang Liu[1], Fei Li[1,✉], Shiyu Zhou[1], Jianfei Hua[1], Warren B. Mori[2], Chan Joshi[2] and Wei Lu[1,✉].

[1]Department of Engineering Physics, Tsinghua University, Beijing 100084, China

[2]University of California Los Angeles, Los Angeles, California 90095, USA

The emergence of multi-petawatt laser facilities is expected to push forward the maximum energy gain that can be achieved in a single stage of a LWFA to tens of GeV, which begs the question - is it likely to impact particle physics by providing a truly compact particle collider? Colliders have very stringent requirements on beam energy, acceleration efficiency and beam quality. In this article, we propose a LWFA scheme that can for the first time simultaneously achieve hitherto unrealized acceleration efficiency from the laser to the electron beam of >20% and a sub-one percent energy spread using a stepwise plasma structure and a nonlinearly chirped laser pulse. Three-dimensional high-fidelity simulations show that the nonlinear chirp can effectively mitigate the laser waveform distortion and lengthen the acceleration distance. This combined with an inter-stage rephasing process in the stepwise plasma can triple the beam energy gain compared to that in a uniform plasma for a fixed laser energy thereby dramatically increasing the efficiency. A dynamic beam loading effect can almost perfectly cancel the energy chirp that arises during the acceleration, leading to the sub-percent energy spread. This scheme is highly scalable and can be applied to peta-watt LWFA scenarios. Scaling laws are obtained that suggest electron beams with energy gain of >100 GeV, charge of 2 nC, and with an energy spread <1% can be realized with a high laser pulse to particle beam energy transfer efficiency in a LWFA driven by a peta-watt laser, which could be the basis for a proof of concept of one arm of a future electron-positron collider.

In the past four decades the field of laser wakefield acceleration (LWFA) has witnessed numerous milestones[1-8]. The extremely high acceleration gradient and microscopic electromagnetic field structure of a LWFA can provide compact and cost-effective accelerators for high-energy physics research, advanced light sources, imaging using electrons and radiation[9,10] and many other research and practical applications. Emerging multi-peta-watt laser projects worldwide and the expected increases in repetition rate and thereby average power of these lasers will provide a further impetus for LWFA to achieve its ultimate scientific goal, a next-generation high-energy particle collider. The principal challenges in realizing even a proof-of-concept LWFA-based compact collider are the simultaneous demands for > 100 GeV beam energy, tens of percent of energy transfer efficiency from the drive laser pulse to the accelerating beam, and ultra-high beam quality (ultra-low transverse emittance, < 1% r.m.s. energy spread, and nano-Coulumb of charge). Despite the fact that impressive progress has been made in LWFA experiments thus far, these advances have invariably focused on maximizing the highest energy (presently at <10 GeV) which is a necessary but not sufficient requirement to build a future collider based on this paradigm changing technology. An accelerator based on such limited energy gain in a single stage necessitates multi-stage operation to reach energies of interest to particle physics and energy recovery of the unspent laser beam. There are of course other challenges such as generating spin polarized beams and an overall rep-rate needed to achieve the necessary luminosity in the collision (interaction) volume. Furthermore, similar considerations need to be addressed for the positron arm of a future e-e+ collider. While a comprehensive design of a LWFA-based e-e+ collider is beyond the scope of any single article, it is possible to address issues for the electron arm. In this article - using particle-in-cell (PIC) code simulations and scaling laws - we show that it is possible to design a single 100 GeV high-efficiency LWFA stage that can provide collider-like beam quality using PW-class lasers that are rapidly coming online around the world. The resulting electron beam would already meet the requirements for the electron arm of a Higgs factory in terms of beam quality, average energy and charge[11,12].

In the beam-driven plasma-based acceleration concept, often referred to as plasma wakefield acceleration (PWFA), energy transfer efficiencies from the drive beam to the trailing beam of 60% have been demonstrated in simulations[13,14] and maximum efficiencies of 30% have been inferred in experiment[8]. However, for a LWFA stage, simulations and experiments have not demonstrated an energy transfer efficiency from the laser pulse to the accelerated electron beam beyond a few percent. Such low acceleration efficiencies of LWFAs primarily stem from the laser pulse shape distortion that occurs because of group velocity dispersion that downshifts frequency of the photons (photon deceleration) and dephasing between the accelerating beam and the continuously temporally evolving laser pulse. Specifically, the non-uniform density and relativistic mass of background electrons in a plasma wake make the photon deceleration rate vary along the laser, leading to longitudinal and transverse pulse distortion before significant pump depletion can occur. Second, the group velocity of the evolving laser and the resulting wake's phase velocity are lower than the speed of relativistic electron bunches. Therefore, there is significant phase slippage between accelerated particles and the wake well before pump depletion of the driver occurs. Moreover, the accelerated beam often has a large correlated energy spread (energy chirp) due to the non-negligible bunch length relative to the (mm-scale) accelerating field structure. Methods have been proposed to reduce the energy spread[15-20], mitigate dephasing[21-29] and pulse depletion[30-32], but a comprehensive proposal is still lacking to simultaneously achieve high overall efficiency, low energy spread and emittance preservation of a LWFA stage.

In this article, we propose a novel scheme that combines a nonlinearly chirped driving laser pulse with a tri-plateau density structure. By using a nonlinearly chirped laser pulse, the axial laser distortion can be significantly mitigated over much longer propagation distance[33,34], leading to a more stable LWFA stage with high laser to wake conversion efficiency. The tri-plateau density structure has three axially uniform sections of progressively higher density connected by upramps, combining with radially parabolic plasma channels (see Fig.1). The use of such a properly designed plasma structure can significantly mitigate the dephasing effect and thereby increase the energy gain and overall efficiency. These two ideas together lead to energy gains three times larger compared with a single uniform density LWFA. At the same time, we discovered that a dynamic beam loading (DBL) effect[35], where the loaded wake initially induces a correlated energy spread that can be then naturally removed as the beam loading effect changes during the wake propagation, leads to accelerated beams with extremely low energy spreads (<1%). The cumulative DBL effect is fully controllable and the optimal output energy spread can be achieved by tuning the density and length of each density plateau. Furthermore, simulations have confirmed that the proposed scheme is highly scalable. For tens to hundreds of TW drive laser scenarios, the scheme can output 0.6~10 GeV electron beams with equally high energy efficiency and low energy spread. Scaling the scheme to the PW-laser-driven regime is expected to result in a remarkable 100-GeV energy gain of a 2nC electron beam.

We rely on high fidelity simulations using OSIRIS to illustrate this high laser coupling efficiency, low energy spread concept. In [Fig .1(a)], a 125pC witness electron beam is accelerated through a tri-plateau density profile to ~600MeV in a wake driven by a laser pulse with only 340 mJ energy. The resulting energy transfer efficiency from the laser to the electron beam is >21%. High efficiency is achieved for relatively low laser normalized vector potential $a_0 \equiv \frac{eA_0}{m_e c^2}$ in the blowout regime[36,37]. According to the scaling law[38], the beam energy gain is $E_b \propto a_0^{5/2}$ and the laser pulse energy is $E_l \propto a_0^{7/2}$ (assuming a round laser pulse $c\tau \sim w_0$), and the energy transfer efficiency is thus $\Gamma \sim \frac{E_b}{E_l} \sim a_0^{-1}$. Here, we chose $a_0 = 1.67$ to ensure the excitation of plasma wake in the blowout regime and a relatively high energy transfer efficiency. The resulting laser power is not sufficient for self-guiding[39], so a tri-plateau plasma structure with a parabolic transverse density of the form $n_p(r) = n_{p0}\left(1 + \frac{\Delta n r^2}{w_0^2}\right)$ is used where $\Delta n = 0.5$ such that the laser with a spot size $w_0$ is free of diffraction throughout the 6-mm-long plasma. Because the density of each plateau increases in discrete steps [Fig. 1(b)], the size of the wake cavity shrinks as it transits from one stage to the next stage [Fig. 1(a)]. The on-axis densities, $n_{p0}$, of the three plateaus are $2 \times 10^{18}$ cm$^{-3}$, $3.13 \times 10^{18}$ cm$^{-3}$ and $8.3 \times 10^{18}$ cm$^{-3}$, and the corresponding lengths of plateau are 1.73 mm, 2.67 mm, 0.15 mm, respectively. Adjacent plateaus are connected by a linear up-ramp of 0.5 mm, and the slopes at both the entrance and the exit for the first and last plateau are 0.25 mm long. Most of the energy gain occurs in the uniform density plateau regions - in this sense the exact lengths of the density upramps between the stages is not critical. The laser pulse has a Gaussian transverse profile, focused to a beam waist $w_0 = 10.7$ μm at the midpoint of the density upramp of the first plateau, and a sin$^2$-shaped temporal envelope of the field with a 21.5 fs FWHM pulse length. The laser pulse is nonlinearly chirped to partially compensate the nonlinear dispersion (see Methods). For ease in interpretation a bi-flattop electron beam with 3.76 μm beam length is initialized behind the drive laser with an initial energy of 51.1 MeV ($\gamma = 100$) and 0.1% energy spread. The resulting beam energy is insensitive to the specific longitudinal profile for a Gaussian-like laser pulse, and >21% energy efficiency with <1% energy spread can be achieved. The initial emittance is 0.1 mm·mrad and the beam is focused to the midpoint of the first upramp with a spot size of 0.5 μm. Compared with the uniform [the yellow line in Fig. 1(b)] and bi-plateau [the red lines in Fig. 1(b)] channel structure, the final energy gain of the tri-plateau structure is almost tripled, significantly increasing the acceleration efficiency. The length of the uniform plasma is 8 mm and the lengths of the two plateaus for the bi-plateau are 1.73 mm and 2.92 mm respectively. Numerical parameters are given in the methods sections.

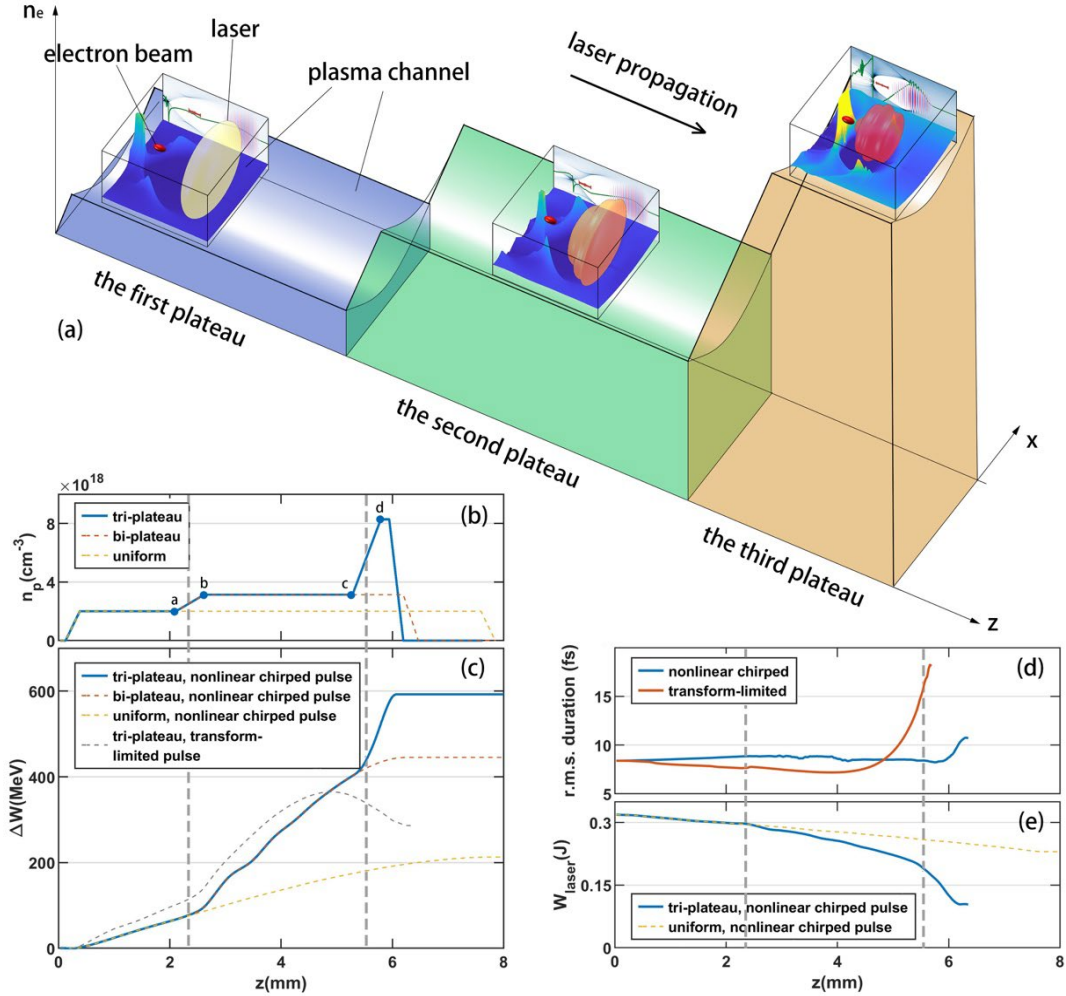

**Figure 1**. (a) Schematic of the high-efficiency LWFA scheme in the tri-plateau plasma channel. (b) On-axis density distribution of the tri-plateau structure. (c) Averaged energy gain in different cases. (d) Pulse length (rms) evolution of the nonlinear chirped pulse (blue line) and the transform-limited pulse (red line) in the tri-plateau structure. (e) Energy consumption of the nonlinear chirped laser pulse in a uniform plasma channel and the tri-plateau structure.

Because ultra-relativistic electrons essentially move at the speed of light, the accelerated e-beam will gradually outrun the accelerating phase of the plasma wake excited by lasers (dephasing). This causes the acceleration gradient felt by the electron beam to decrease or even become negative as the laser propagates in a uniform plasma. Since the dephasing length is typically much shorter than the pump depletion length, the acceleration gradient will be significantly reduced far before the laser energy is depleted, thus greatly limiting the acceleration efficiency. Due to this dephasing process, the spatial gradient of the energy gain in the laboratory frame gradually reduces and appears parabolic as depicted by the yellow dashed in Fig. 1(c). On the other hand, during the acceleration process, the laser loses energy in a nearly linear manner as shown by the yellow dashed in Fig. 1(e) and eventually only 1/3 of the initial energy is transferred to the plasma wake. In our tri-plateau scheme, a rephasing process occurs both in the density uplift and the higher density plateau that follows it. In the rephasing process, the wavelength of the plasma wake shrinks due to the increase of the plasma density and e-beam finds itself once again close to the maximum acceleration phase, significantly improving the acceleration efficiency. Since the acceleration gradient also increases with ~ square root of the density, the slope of the gain curve (blue line) increases sharply from one plateau to the next. As shown in Fig. 1(c). The energy gain of each stage is insensitive to the plasma density albeit the acceleration distance (being proportional to the dephasing length) becomes shorter as the density increases. Using the scaling laws in the blowout regime [38], for the accelerating field, $E_z \propto n_p^{1/2} a_0^{1/2}$, and the dephasing length, $L_d \propto n_p^{-1}$, the energy gain scales as $\Delta E \propto n_p^{-1/2} a_0^{1/2}$. Due to the focusing and steepening of the laser pulse, $a_0$ increases so that $\Delta E$ does not change much as $n_p$ increases. As a result, the energy gain is nearly tripled compared with that of the uniform plasma case.

As shown by the blue solid line in Fig. 1(c), the energy gain curve now increases almost linearly in each stage and hence a high beam loading efficiency is maintained continuously for the 125pC beam charge. The beam loading efficiency (wake-to-beam efficiency), defined as the ratio of energy extracted by the particles from the wake to the energy cost of the laser, is calculated by $N\langle\Delta W\rangle/(W_i - W_o)$ in each plateau, where $N$ is the electron number of the beam, $\langle\Delta W\rangle$ is the averaged energy gain of the electron beam in this plateau and $W_i/W_o$ is the laser energy entering/exiting this plateau. In the first two plateaus, the wake-to-beam efficiency is 49.6% and then drops to 22.3% in the third plateau. More laser energy is utilized in the tri-plateau plasma; 67.5% of the laser energy is consumed in the proposed scheme while only 34.0% is consumed in a single uniform plasma channel. The total efficiency is the product of how much of the laser energy is transferred to the wake (roughly the laser energy utilization percentage) times the beam loading efficiency, which is > 30% in this case. As a result, a total efficiency of > 21% is obtained which is the highest value achieved to date for a fully self-consistent LWFA concept.

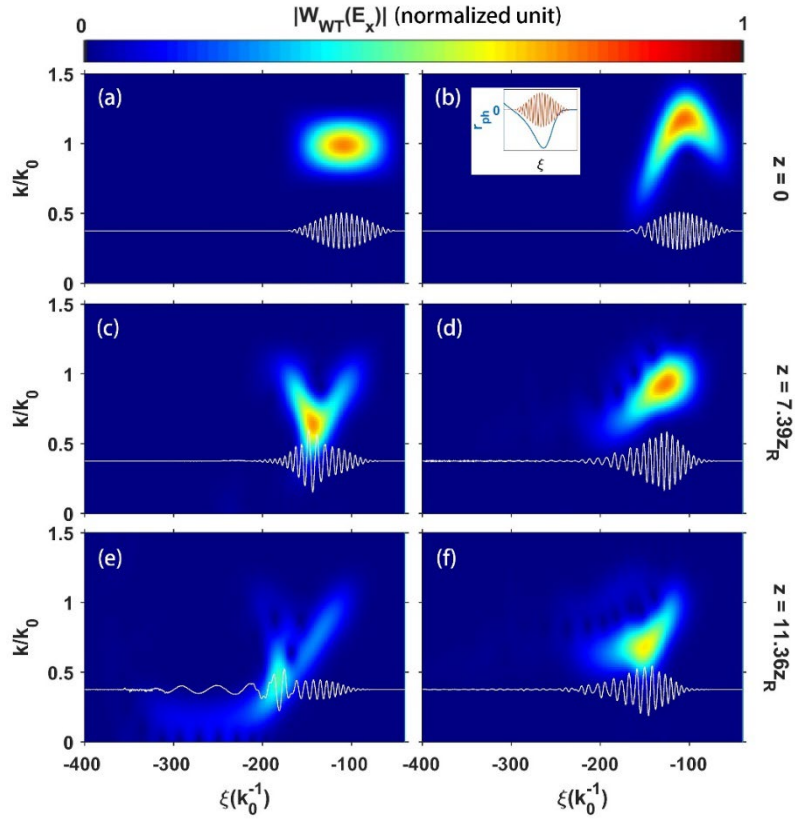

**Figure 2**. On-axis spectrograms (in color-obtained by Wigner-wavelet transforming the electric field) of the laser pulse at various times for an initially transform-limited laser pulse (a, c and e) and a chirped laser pulse (b, d and f) propagating through a single plateau. (a), (c) and (e) show the spectrograms of the original unchirped or transform limited laser pulse at $z = 0$, $z = 7.39z_R$ and $z = 11.36z_R$, where the Rayleigh length $z_R$ of the original laser pulse is 447 μm. The inset in (b) shows the laser field $E_x$ by the red line and the frequency shift rate by the blue line. As a comparison, (b), (d) and (f) show the spectrograms of the initially chirped laser pulse at the same distances. The spectrograms in all the subplots are obtained by carrying out wavelet transform of $E_x$, i.e., $|W_{WT}(E_x)| = \left|\int_{\xi-\frac{\delta}{2}}^{\xi+\frac{\delta}{2}} E_x(\xi')e^{ik\xi'}d\xi'\right|$ with the width of the transform window $\delta = 25.4$μm. The white lines are the on-axis lineouts of $E_x$. The initially chirped laser pulse is able to maintain an overall similar pulse envelope shape (f) as it has at the beginning (b) whereas the initially unchirped pulse spreads rapidly due to photon deceleration and group velocity dispersion.

We note that in this proposed scheme the laser energy is efficiently utilized because the pump laser pulse has a nonlinear chirp with frequency increasing from front to middle and then decreasing from the middle to the back, allowing the pump waveform to self-compensate for photon deceleration due to wake formation and maintain its shape. The nonlinear frequency chirp is of the form $\frac{k(\xi)}{k_0} = c_0 + c_1\xi + c_2\xi^2$, where $k(\xi)$ and $k_0$ are the local and central wavenumber and $\xi \equiv ct - z$. The chirp coefficients $c_0$, $c_1$, $c_2$ are 1.2, $1.79 \times 10^{-3}k_0$ and $-9.7 \times 10^{-5}k_0^2$ respectively. Such a nonlinear chirp can either be realized by nonlinear cross-phase modulation[40] or nonlinear pulse compression by customized gratings[41,42]. In the blowout regime[36,37], the driving laser pulse with the optimal pulse length resides at the first half of the ion cavity where the photon deceleration rate $r_{ph} < 0$ (see Methods). The occurrence and effects on pulse distortion of photon deceleration/pump depletion are illustrated in Fig. 2. The left column frames are taken from a simulation with a transform limited pulse while the right column is for a laser with a nonlinear frequency chirp. For these examples a laser with identical parameters as above is sent through a single plateau with a $n_{p0} = 3 \times 10^{18} \text{cm}^{-3}$ and $\Delta n = 0.5$. For each case, the axial lineout and spectrogram of the laser are shown at propagation distances, 0 mm, 3.81 mm and 5.08 mm respectively. Initially the middle of the pulse experiences a more intense photon deceleration than the head and tail of the pulse [see the inset in Fig. 2(a)]. This can be seen from Fig. 2(a), 2(c) and 2(e), where the on-axis spectrogram for the transform-limited laser evolves into a V-shape, indicating as expected that the middle part suffers from a faster photon deceleration. The pulse is significantly lengthened as a result of the resulting envelope distortion [Fig. 2(e)]. On the other hand, as shown in Fig 2(b), (d) and (f), the initial nonlinear chirp nearly perfectly balances the redshifting, and the waveform is well preserved throughout the simulation. Off-axis evolution of the spectrogram has a similar pattern shown by Fig. 2, which is supplied in Methods.

Such a pulse can propagate much further than the pump depletion limit of a transform-limited pulse of similar pulse width and peak intensity. The importance of this chirp is seen in the grey line of Fig 1(c) where it can be seen that no energy gain arises in the third stage without the chirp (transform limited pulse). The reason for saturation of the energy gain can be seen in Fig. 1(d) to be that the pulse length increases rapidly for the laser pulse that has no chirp in the final stage. In addition to the high efficiency, the beam energy spread of <1% can be obtained owing to the DBL. The mechanism of how the DBL effect maintains a low energy spread will be elaborated on in the later section.

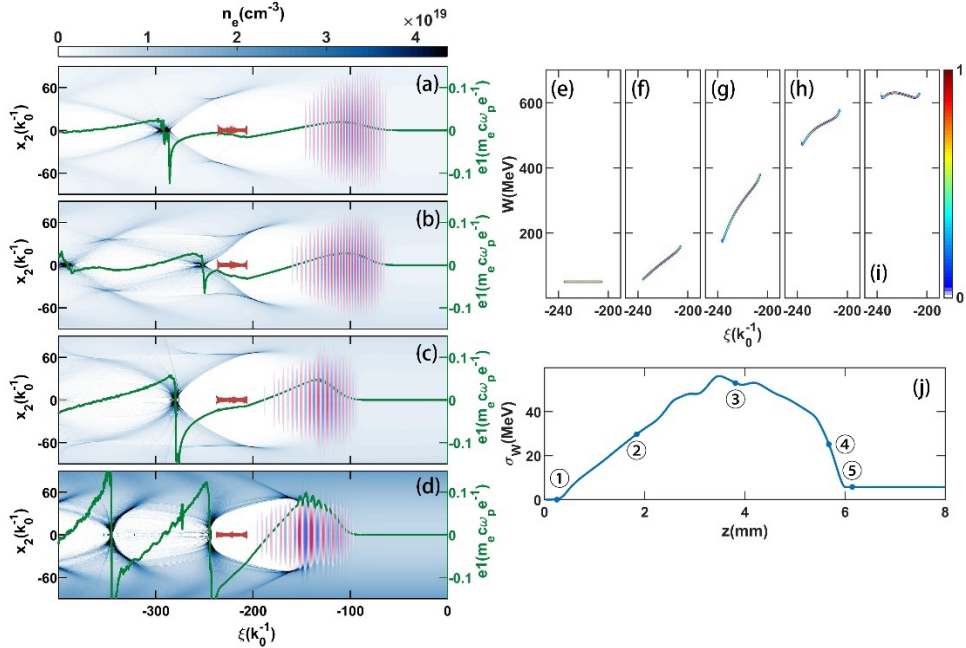

**Figure 3.** (a)-(d) Snapshots of the density and the electric field distributions before/after the second [(f) and (g)] and third plateau [(h) and (i)], respectively. (e)-(i) Snapshots of beam longitudinal phase space corresponding to the tagged points in (j). (j) Absolute beam energy spread through the tri-plateau

structure.

The use of stepwise plasma density plateaus connected with density up-ramps not only significantly increases the beam energy gain, but also is capable of accelerating a large amount of beam charge with an extremely small energy spread through a dynamic beam loading (DBL) effect. The beam loading effect can be understood by tracking the accelerating field felt by the beam[13]. In Figs. 3(a) to 3(d), the laser field, lineout of the accelerating field (green line), plasma, and beam density are plotted at the start to end points of each transition section (a to d) in Fig. 1(b). In Figs. 3 (e)-(i) the beam energy vs. axial position is shown at the same propagation distances. It can be seen that the plasma wake is over-loaded ($\frac{dE_z}{d\xi} < 0$) during the first and at the entrance of the second plateau [Figs. 3(a) and (b)], leading to a negative energy chirp in the electron beam. On the other hand, in the third plateau, and during the exit of the second and third plateaus the wake is under-loaded ($\frac{dE_z}{d\xi} > 0$) [Fig. 3(c) and (d)] such that the energy chirp caused by the first plateau is almost perfectly removed. The transition from an over- to under-loaded wake happens in the second plateau [between Figs. 3(b) and (c)].

This mechanism can be qualitatively described by the nonlinear theory of the blowout regime[36]. We find that the gradient of $E_z$ (see Methods) can be expressed as

$$\frac{1}{E_p}\frac{dE_z}{d\xi} \approx -\frac{2\lambda}{k_p r_b^2} + \frac{k_p}{2}, \qquad (1)$$

where $\lambda = k_p^2 \int_0^{+\infty} \frac{n_b}{n_p} r\,dr$ is the normalized linear charge density of the beam, $E_p \equiv \frac{m_e c \omega_p}{e}$ is the characteristic electric field in the plasma, and $r_b$ is the blowout radius. The linear charge density $\lambda$ which is independent of plasma density can be assumed to not depend on the propagation distance. However, the value of $r_b$ at the location of the beam changes (increases) from dephasing within an individual plateau, and $k_p$ increases when crossing into the next plateau. For the parameters simulated, during the first plateau the dominant term is $-\frac{2\lambda}{k_p r_b^2}$ while it becomes $\frac{k_p}{2}$ in the third plateau. In the second plateau, the dominant term changes from $-\frac{2\lambda}{k_p r_b^2}$ to $\frac{k_p}{2}$ as $r_b$ increases because of the increased plasma density and dephasing. The final energy chirp is obtained by integrating Eq. (3) with respect to the propagation distance, and eventually vanishes at the optimal distance [Fig. 3(i)]. As seen Figs 3(e)-(i), the beam energy chirp increases first and then decreases, and the residual energy chirp is almost fully compensated, which agrees with the analysis above. The final relative energy spread reaches as low as 0.83% (FWHM) while a considerable beam charge of 125pC is accelerated.

Assuming a fixed $a_0$ and $\lambda_0$, and that the laser is guided by a channel, the key physics can be viewed approximately self-similar, and scaling laws can be straightforwardly obtained (see Methods). When the plasma density changes from $n_p$ to $n'_p$, the characteristic lengths of the key physics can be scaled by a factor $F = \sqrt{n_p/n'_p}$. If we proportionally scale the focal waist and pulse length of the laser through $w'_0 = F \cdot w_0$ and $\tau' = F \cdot \tau$, the acceleration gradient, maximum beam energy gain and the number of electrons that can be accelerated scale as, $E_z' = F^{-1} E_z$, $\langle \Delta W \rangle' =$

$F^2 \cdot \langle \Delta W \rangle$ and $N' = F \cdot N$ respectively according to the scaling law of LWFAs [38] (see the Methods for the derivation). The energy transfer efficiency $\eta'$ is an invariant under this scaling since

$$\eta' = \frac{N' \cdot \langle \Delta W \rangle'}{W_{laser}'} = \frac{FN \cdot F^2 \langle \Delta W \rangle}{F^3 W_{laser}} = \eta. \qquad (2)$$

We carried out simulations with $F = 2$ and $F = 4$ to validate the scaling law, and the results are summarized in table 1. In these simulations, the initial parameters of laser, plasma and electron beam are scaled from case 1 and the results (energy gain, efficiency) show excellent agreement with what the scaling law predicts as shown in Fig. 4(b). In each case the final energy spread remains below 1%. Case 3 corresponds to a 500pC electron beam being accelerated to ~10 GeV by a 22-J laser pulse which is within the capability of modern PW lasers. The calculated energy transfer efficiency greatly exceeds that obtained in previous LWFA experiments and that described in previous simulations. Experimental verification of case 3 can be attempted in the near term. Furthermore, we can extrapolate the proposed scheme to $F = 16$ in column 4. In this case, an LWFA driven by a 1.4kJ, 340fs laser pulse would generate a 2nC electron beam with ~150GeV energy with an average acceleration gradient of 6 GeV/m. Laser with such energy and beams with such charge are already available. The proposed scheme could thus provide the electron arm of a Higgs factory, with ~20% efficiency and sub-1% energy spread. For comparison, a scaled case (Case 5) using a uniform plasma channel is also listed in Table 2. In this case, only about one third of the energy gain in Case 4 is achieved, and the energy spread grows up to 15 %.

**Table 2**. Parameter designs of Case 1 to Case 4 for $a_0 = 2$ and $\lambda_0 = 800$ nm according to the scaling law, in which case 1~3 are PIC simulation results and case 4 is an extrapolation to obtain the 145.9 GeV electron beam. As a comparison, Case 5 uses the same laser pulse as Case 4 but in a matched uniform plasma channel, which is scaled from the case of the yellow dashed in Fig. 1(c).

| Case | 1 | 2 | 3 | 4 | 5 (uniform channel) |
|---|---|---|---|---|---|
| Scaling factor $F$ | 1 | 2 | 4 | 16 | 16 |
| Averaged energy gain (GeV) | 0.57 | 2.28 | 9.88 | 145.9 | 51 |
| Laser peak power (TW) | 15.3 | 61.2 | 245 | 3917 | 3917 |
| Pulse length (fs) | 21.5 | 43.0 | 86.0 | 344 | 344 |
| $w_0$ (μm) | 10.7 | 21.3 | 43.0 | 171 | 171 |
| Laser pulse energy (J) | 0.34 | 2.76 | 22.1 | 1393 | 1393 |
| Plasma density of first plateau (cm$^{-3}$) | 2e18 | 5e17 | 1.25e17 | 7.81e15 | 7.81e15 |
| Plasma length (mm) | 6.1 | 48.8 | 390 | 2.5e4 | 3.3e4 |
| Electron beam charge (pC) | 125 | 250 | 492 | 2000 | 2000 |
| Energy spread (FWHM) | 0.42% | 0.65% | 0.63% | <1% | ~15% |
| Energy efficiency | 21.1% | 21.2% | 21.1% | ~21.1% | ~7.4% |

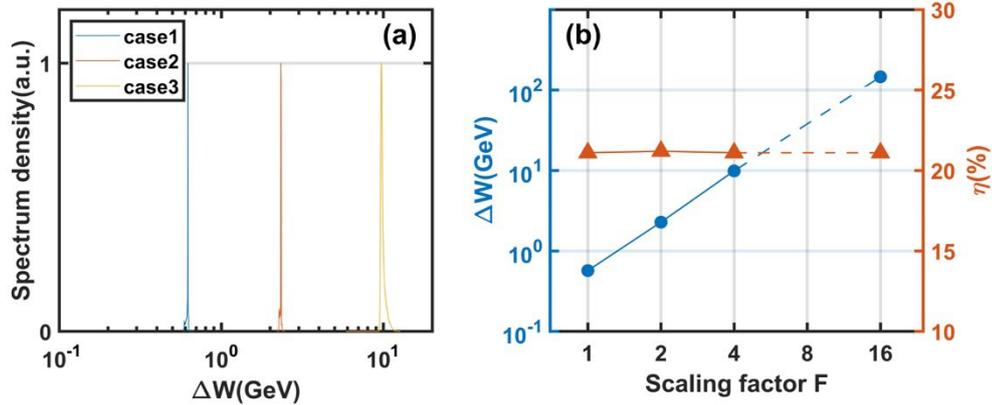

**Figure 4**. (a) The energy spectra and (b) the averaged beam energy gain and transfer efficiency of Case 1 to 3 in Table 2.

## Methods

**High fidelity PIC simulation.** The propagation and the evolution of the laser pulse is of crucial importance in LWFA research. Thus high fidelity modeling of the laser-plasma interaction including nonlinear dispersion effects is critical for the calculation of energy gain and energy transfer efficiency. Standard PIC simulations utilize a second order Yee solver to solve the EM fields which could suffer from significant numerical dispersion, numerical Cerenkov radiation (NCR), and self-fields from the beam for the propagation distances considered here. These numerical errors may cause the PIC simulations to overestimate the dephasing effect that is closely related to the energy efficiency from laser pulse to the electron beam, and the growth of the emittance and energy spread of the beams. In this article, a customized finite difference EM solver which can eliminate the numerical dispersion, NCR, and self-forces is adopted[43-46].

All the simulations are conducted using the quasi-3D version of the PIC code OSIRIS. For the $F = 1$ case the simulation window has a dimension of $150k_0^{-1} \times 400k_0^{-1}$ with $500 \times 2000$ cells in the $r$ and $z$ directions, respectively. This corresponds to cell sizes of $\Delta r = 0.30k_0^{-1}$ and $\Delta z = 0.20k_0^{-1}$ (where the wave number $k_0 = 2\pi/\lambda_0$ and the wavelength $\lambda_0 = 800$ nm). The time step is chosen as $\Delta t = 0.1c^{-1}k_0^{-1}$. In the scaled cases of $F = 2$ and $F = 4$, the sizes of the simulation window and transverse cell size are scaled accordingly while the longitudinal cell size is kept at $\Delta z = 0.20k_0^{-1}$. There are 10 macro particles in each cell and 8 duplications in theta. Only the m=0 and m=1 azimuthal modes are kept in the simulations. In the quasi-3D geometry, the laser field only has m=1 azimuthal Fourier harmonics and the symmetric physics of the plasma wake can be fully captured by the m=0 harmonics. Therefore, we took advantage of this feature of the quasi-3D code to clearly separate the laser field from the plasma wake in the calculation of energy transfer efficiency.

**Waveform evolution of the nonlinearly chirped laser pulse.** Due to the self-phase modulation of the laser pulse in the plasma, the laser pulse loses its energy by the frequency downshifting (photon deceleration) with the rate expressed as[47-49]:

$$r_{ph} \equiv \frac{1}{k}\frac{\partial k}{\partial t}, \tag{3}$$

where $k$ is the local wavenumber of the laser. This photon deceleration rate has a distribution along $\xi$ [the inset of Fig. 2(b)], and is almost non-evolving during most of its propagation length. In the proposed high efficiency scheme, the nonlinear chirp is added to the laser pulse to offset this photon deceleration. Another option is to shape the pulse to provide a wake with a constant photon deceleration rate throughout the entire pulse[50].

In addition to the on-axis waveform evolution of the nonlinearly chirped pulse (Fig. 2), off-axis waveform evolutions with different transverse position up to 11.1μm are provided in Fig. 5 for integrality, noting that the matched $w_0 = 10.7$ μm. Each row shows the laser time-frequency distribution at a different time when the laser pulse propagates 0, $7.39z_R$ and $11.36z_R$ with $z_R = 447$ μm is the Rayleigh length, while each column shows the spectrograms with different transverse positions at $x_2 = 3.7$μm, 7.4μm and 11.1μm, noting that the beam waist of the initial laser pulse $w_0 = 10.7$ μm. The off-axis waveforms show similar evolution with the on-axis waveform [Fig. 2(b), (d) and (f)].

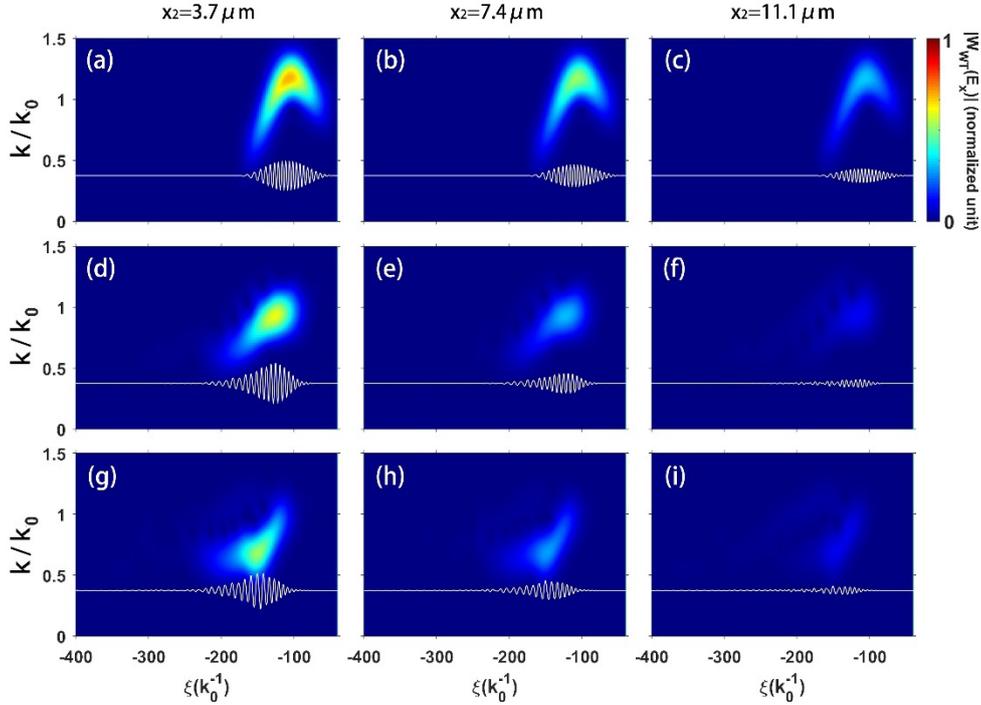

**Figure 5**. Off-axis spectrograms (in color-obtained by Wigner-wavelet transforming the electric field) of the nonlinearly chirped laser pulse at various times propagating through a single plateau with different transverse positions. The first, second and third rows show the spectrograms at $z = 0$, $z = 7.39z_R$ and $z = 11.36z_R$, where the Rayleigh length $z_R$ of the original laser pulse is 447 μm. The different columns show the spectrograms for different transverse positions with $x_2 = 3.7$ μm, 7.4μm and 11.1μm. Other parameters of the wavelet transform are same with Fig. 2. The white lines are the corresponding lineouts of laser transverse electric field.

**Calculation of the gradient of $E_z$.** To derive the expression of $E_z$, the nonlinear theory of the bubble regime[36] is followed here. In the ultra-relativistic limit where the maximum blowout radius $k_p R_b \gg 1$, the trajectory of the inner-most particle is given by

$$r_b \frac{d^2 r_b}{d\xi^2} + 2\left(\frac{dr_b}{d\xi}\right)^2 + 1 = \frac{4\lambda(\xi)}{k_p^2 r_b^2}, \tag{4}$$

where $r_b(\xi)$ is the radial position of the inner-most particle in the bubble sheath, and $\lambda = k_p^2 \int_0^{+\infty} n_b/n_p r dr$ is the normalized linear charge density of the electron bunch. The longitudinal electric field $E_z$ can be expressed as a function of $r_b$[38],

$$\frac{E_z}{E_p} = \frac{1}{E_p}\frac{\partial \psi}{\partial \xi} \approx -\frac{k_p}{2} r_b \frac{dr_b}{d\xi}, \tag{5}$$

where $\psi \equiv \phi - A_z/c$ is the pseudo-potential of the plasma wake. The gradient of the accelerating field $dE_z/d\xi$ is obtained via differential of Eq. (5),

$$\frac{1}{E_p}\frac{dE_z}{d\xi} \approx -\frac{k_p}{2}\left(\frac{dr_b}{d\xi}\right)^2 - \frac{k_p}{2} r_b \frac{d^2 r_b}{d\xi^2} = -\frac{2\lambda}{k_p r_b^2} + \frac{k_p}{2} + \frac{k_p}{2}\left(\frac{dr_b}{d\xi}\right)^2$$

$$\approx -\frac{2\lambda}{k_p r_b^2} + \frac{k_p}{2}. \tag{6}$$

In equation (6), the term $\frac{k_p}{2}\left(\frac{dr_b}{d\xi}\right)^2$ is omitted since it is negligible for most part of the bubble.

**Scaling of the simulations.** The initial parameters for the scaled simulations are guided by LWFA scaling laws[38], starting from the scaling of the accelerating wakefield structure. For convenience, real and normalized units are used in this section. Through simulations it has been found that a relatively stable propagation of the ion cavity is realized when $k_p w_0 \simeq k_p R_b \simeq 2\sqrt{a_0}$, where $w_0$ is the initial laser beam waist, and $R_b$ is the blowout radius of the ion cavity[36,37]. For the parameters being considered here, these relationships hold in a parabolic density channel if the density at the bottom of the channel is used. The connection between the plasma density and the size of the ion cavity is established as $n_p \propto F^{-2}$ because $k_p \propto \sqrt{n_p}$, where $F$ is the scaling factor of the radius of the ion cavity. To efficiently drive the wake, the laser pulse length should fill the first half of the ion cavity, thus $c\tau_{\text{FWHM}} \propto R_b$, where $\tau_{\text{FWHM}}$ is the pulse duration. Hence the laser pulse energy scales with $F^3$. The length of the plasma structure $L_{acc}$ is characterized by the dephasing length $L_\phi = \frac{4}{3}\frac{k_0^2}{k_p^2}\frac{\sqrt{a_0}}{k_p}$, providing $L_{acc} \propto F^3$. Using the scaling law of the accelerating field, $E_z \propto \frac{m_e c \omega_p}{e} \propto F$, the expected beam energy gain is obtained as $\langle \Delta W \rangle \propto F^2$. Another initial parameter to be determined is the loaded electron number $N$, and based on [13,36,51] it should scale with the electron number that is expelled from the ion cavity, leading to $N \propto n_p R_b^3 \propto F$. The relevant formulas and scaling factors are summarized in Table 2.

As for the PIC simulations, the simulation window should scale with the ion cavity, e.g. $r_{\text{window}} \propto F$ and $z_{\text{window}} \propto F$. The transverse cell size $\Delta r$ scales with $R_b$ to resolve the plasma wavelength while the longitudinal cell size $\Delta z$ is fixed to resolve the laser wavelength.

Table 2. Scaling of the laser, plasma and beam parameters with scaling factor $F$ for fixed $a_0$ and $\lambda_0$.

| | Formula | Scaled parameters for fixed $a_0$, $\lambda_0$ with scaling factor F |
|---|---|---|
| Laser | $k_p w_0 = 2\sqrt{a_0}$ | $w_0' = F \cdot w_0$ |
| | $c\tau = w_0$ | $\tau' = F \cdot \tau$ |
| Plasma | $k_p^{-1} \propto 1/\sqrt{n_p}$ | $n_p' = F^{-2} \cdot n_p$ |
| | $k_p R = 2\sqrt{a_0}$ | $R' = F \cdot R$ |
| | $L_\phi = \frac{3}{4}\frac{k_0^2}{k_p^2}\frac{\sqrt{a_0}}{k_p}$ | $L_\phi' = F^3 \cdot L_\phi$ |
| Beam | $\langle \Delta W \rangle = \frac{2}{3} m_e c^2 \frac{k_0^2}{k_p^2} a_0$ | $\langle \Delta W \rangle' = F^2 \cdot \langle \Delta W \rangle$ |
| | $N = \frac{8}{15 k_0 r_e} \sqrt{\frac{P}{m_e^2 c^5 / e^2}}$ | $N' = F \cdot N$ |
| | $\sigma_z \propto R$ | $\sigma_z' = F \cdot \sigma_z$ |

## Data availability

The data that support the findings of this study are available from the corresponding author on reasonable request.

## Code availability

The codes that support the findings of this study are available from the corresponding authors upon reasonable request.

## Acknowledgement

This work is supported by the National Natural Science Foundation of China (NSFC) Grants (No.

11991071, No. 11775125 and No. 11875175), CAS Center for Excellence in Particle Physics. C. Joshi and W. B. Mori. were supported by DOE grant DE-xxxxxxxxx etc.

## Author contributions

W.L. conceived and supervised the project. S.L. and W.L. proposed the main ideals. S.L. and F.L. carried out the simulations. W.L., F.L., S.L., W.M. and C.J wrote the paper. All the authors contributed extensively to the work presented in this paper.

## Competing interests

The authors declare no competing interests.